\documentclass[showpacs,twocolumn,floatfix,preprintnumbers]{revtex4}
\usepackage{amsmath,amssymb}
\usepackage{graphicx}

\usepackage{dcolumn}
\usepackage{bm}
\begin{document}
\title{Cosmic Ray Spectrum and Tachyonic Neutrino}
\author{Guang-Jiong Ni}
        \email{pdx01018@pdx.edu}
        \affiliation{Department of Physics, Fudan University, Shanghai, 200433, China\\
        Department of Physics, Portland State University, Portland, OR 97207 USA}
\begin{abstract}
In the cosmic ray spectrum, there  are two knees (abrupt changes of slopes) located
around the energies $E_{th}=10^{15.5}$ eV and $10^{17.8}$ eV respectively. Based on the
pioneering work by Kostelecky and Ehrlich, we ascribe the first knee to a sudden opening
of the reaction channel $\tilde{\nu}_e+p\rightarrow n+e^+$ when the proton has a velocity
just exceeding a critical value and so can absorb a tachyonic neutrino $\nu_e$ in the
form of an antineutrino $\tilde{\nu}_e$. Similarly, the second knee is triggered by the
reaction $\tilde{\nu}_\mu+p\rightarrow \Lambda+\mu^+$. The fitting of these two values of
$E_{th}$ gives the tachyon mass of $\nu_e$ and $\nu_\mu$ being $0.54$ eV/$c^2$ and $0.48$
eV/$c^2$ respectively, which are in favor of a minimal three-flavor model for tachyonic
neutrino with one parameter $\delta$ being estimated to be $\delta=0.34$ eV.
\end{abstract}

\maketitle

\section{introduction}

It has been known for years that the observed energy spectrum of primary cosmic rays can
be well described by an inverse power law in the energy $E$ from $10^{11}$ to $10^{20}$
eV \cite{1,2}
\begin{equation}\label{eq:1}
  \frac{dJ}{dE}\sim E^{-\gamma},
\end{equation}
where $J$ is the flux in $m^{-2}s^{-1}sr^{-1}$. However, the index $\gamma$ reveals an
abrupt change at around $10^{15.5}$ eV$=3.16\times 10^{15}$ eV$=3.16$ PeV:
\begin{equation}\label{eq:2}
  \gamma=\left\{\begin{array}{lr}2.7,\quad& E\leq 10^{15.5}\;{\rm eV}\\
  3.\quad& E>10^{15.5}\;{\rm eV} \end{array}\right.
\end{equation}
This sudden change in the slope of cosmic ray spectrum (CRS) is usually called the
``knee". It seems that there is a second knee at around $10^{17.8}$ eV$=6.31\times
10^{17}$ eV, then follows the ``ankle" at around $10^{19}$ eV, at which the slope changes
from $\gamma=3.16$ to 2.78 \cite{3}.

Among various tentative explanations for the knee in CRS, the model initiated by
Kostelecky first \cite{4} and then elaborated by Ehrlich \cite{5,6,7} is the most
attractive one. Their basic idea is as follows:

Corresponding to the decay of a neutron
\begin{equation}\label{eq:3}
  n\rightarrow p+e^-+\tilde{\nu}_e
\end{equation}
the ``proton decay"
\begin{equation}\label{eq:4}
  p\rightarrow n+e^++\nu_e
\end{equation}
is considered. For the decay to conserve energy in the proton rest frame [ to be referred
as the $S^\prime$ frame in which an observer Bob (B) is present ], $E_\nu<0$ is needed.
Then the threshold laboratory [ defined by the cosmic background radiation (CBR), to be
referred as the $S$ frame in which another observer Alice (A) is present ] energy for
proton was derived as  ( $\Delta =m_n+m_e-m_p$ ) \cite{5}:
\begin{equation}\label{eq:5}
  E_{th}=\frac{m_p\,\Delta}{|m_{\nu_e}|}=\frac{1.7\times 10^{15}}{|m_{\nu_e}|}\;{\rm eV}
\end{equation}
where $|m_{\nu_e}|=\sqrt{-m^2}$ is the tachyon mass of neutrino. If setting $E$ to be the
energy of knee $\sim 4.5\times 10^{15}$ eV, they found $|m_{\nu_e}|=0.38$ eV/$c^2$.

The idea of how a ``stable" proton can decay was explained by the so-called
``reinterpretation principle" \cite{8,9}. The emitted $\nu_e$ with $E_\nu>0$ in the
process~(\ref{eq:4}) in the $S$ frame could be identified with the process
\begin{equation}\label{eq:6}
  \tilde{\nu}_e+p\rightarrow n+e^+
\end{equation}
with an absorbed $\nu_e$ with $E_{\tilde{\nu}}<0$ from a background sea in the $S^\prime$
frame. Furthermore, the cosmic ray nucleons on their way to Earth would lose their energy
through the chain of decays $p\rightarrow n\rightarrow p\rightarrow n\rightarrow\cdots$,
which not only depletes the spectrum at energies above $E_{th}$ \cite{4} and may also
account for the existence of the ``ankle" and a neutron ``spike" just above $E_{th}$
\cite{5,6}. Moreover, the reason why the GZK cutoff ( cosmic rays with energies much
above $5 \times 10^{19}$ eV were predicted to not reach Earth from distant sources
because of their interaction with photons comprising the CBR) is absent (as shown in
\cite{3}) is because neutrons have a much smaller interaction with the CBR.

The purpose of this paper is to further justify the above model by refining the theory of
tachyon (superluminal particle, i.e., faster-than-light particle) for neutrinos. We will
take both energy and momentum conservation laws into account to rigorously derive the
relation between $E_{th}$ and the tachyon mass m ( real and positive ) of neutrinos. If
setting $E_{th}=3.16\times 10^{15}$ eV, we will find the value of $m$ for $\nu_e$ being
0.54 eV/$c^2$. Moreover, we consider a similar process in the $S^\prime$ frame to create
the neutral hyperon $\Lambda$ (with quark content $uds$ ) and a muon:
\begin{equation}\label{eq:7}
  \tilde{\nu}_\mu+p\rightarrow \Lambda+\mu^+.
\end{equation}
Fitting the energy at the second knee $E^{(2)}_{th}=6.31\times 10^{17}$ eV, we find the
tachyon mass of $\nu_\mu$ around 0.48 eV/$c^2$. The implications of our theory will be
discussed below.

\section{The Lorentz transformation and a solution to superluminal paradox}

\begin{figure}[h]
  \includegraphics{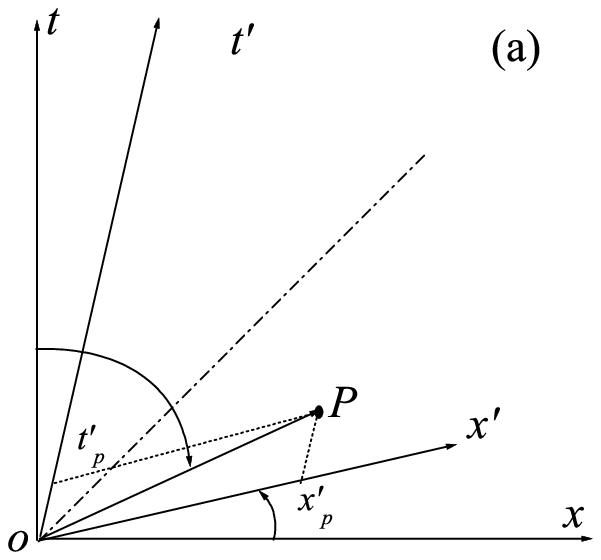}
  \includegraphics{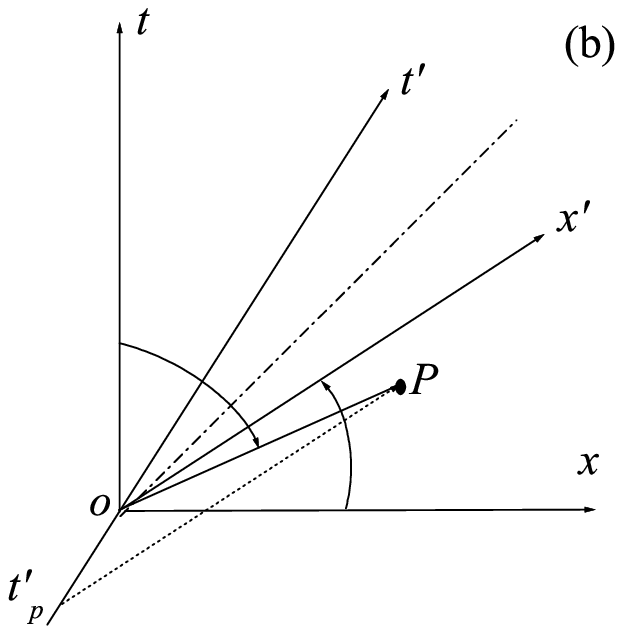}
  \caption{\label{fig:1}A superluminal particle (P) moving along $x$ axis with velocity
  $u>c$. (a) $v<\frac{c^2}{u}, t^\prime_p>0$; (b) $v>\frac{c^2}{u}, t^\prime_p<0$.}
\end{figure}

The reason why many physicists do not believe in tachyons goes back to a strange puzzle
involving tachyon motion. See Fig.~\ref{fig:1} \cite{9,10}. For clarity, we only consider
its motion in a one dimensional space.

A tachyon (P) is moving along the $x$ axis with a velocity $u>c$ in the $S$ frame. Bob
takes another $S^\prime$ frame moving relative to $S$ with velocity $v$. Then if
$v>c^2/u$, the time coordinate of P in the $S^\prime$ frame will become negative:
\begin{equation}\label{eq:8}
   t^\prime<0 \quad (u>c,\; v>c^2/u)
\end{equation}
which was regarded as the ``tachyon traveling backward in time" or ``a violation of
causality" \cite{8}.
\begin{figure}[t]
  \includegraphics{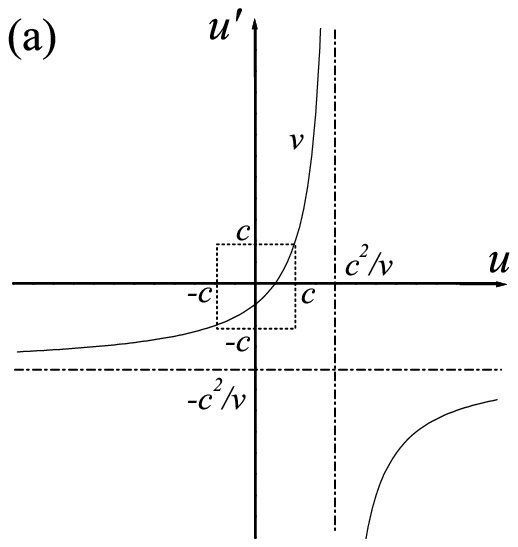}
  \includegraphics{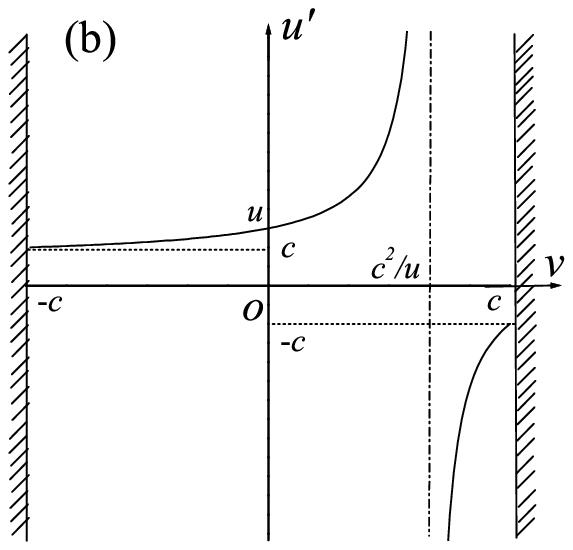}
  \caption{\label{fig:2}Addition of velocities in Lorentz transformation. (a) $u^\prime$ as a
  function of $u$ for a fixed $v$; (b) $u^\prime$ as a function of $v$ for a fixed $u(>c)$}.
\end{figure}

In our opinion, the above puzzle can be better displayed in an alternative way. From the
well known Lorentz transformation (LT), we have the addition law for velocities as:
\begin{equation}\label{eq:9}
    u^\prime=\frac{u-v}{1-uv/c^2}
\end{equation}
where $u^\prime$ is the velocity of tachyon in the $S^\prime$ frame. As shown in the
Fig.~\ref{fig:2} \cite{9,10}, there is a pole at $uv/c^2=1$. For a fixed $u$, when $v$
increases across the singularity $c^2/u$, Bob will see that $u^\prime$ leaps abruptly
from $\infty\rightarrow - \infty$:
\begin{equation}\label{eq:10}
  u^\prime<-c. \quad ( u>c^2/v\;\; {\rm or}\;\; v>c^2/u )
\end{equation}
However, Eq.~(\ref{eq:10}) still remains as a puzzle. According to LT, the momentum
$p^\prime$ and energy $E^\prime$ of tachyon in the $S^\prime$ frame are related to $p$
and $E$ in the $S$ frame as follows:
\begin{equation}\label{eq:11}
  p^\prime=\frac{p-vE/c^2}{\sqrt{1-v^2/c^2}},\quad E^\prime=\frac{E-vp}{\sqrt{1-v^2/c^2}},
\end{equation}
with
\begin{equation}\label{eq:12}
  p=\frac{mu}{\sqrt{u^2/c^2-1}}>0,\quad E=\frac{mc^2}{\sqrt{u^2/c^2-1}}>0.
\end{equation}
Here $m$ is the tachyon mass of a particle with kinematical relation as;
\begin{equation}\label{eq:13}
  E^2=p^2c^2-m^2c^4,\quad u=\frac{dE}{dp}=\frac{p\,c^2}{E}>c.
\end{equation}
Combining (\ref{eq:10}) with (\ref{eq:11}) leads to:
\begin{eqnarray}
  p^\prime\!\!&=&\!\!\frac{m}{\sqrt{1-\frac{v^2}{c^2}}\,\sqrt{\frac{u^2}{c^2}-1}}\,(u-v)>mc>0,\label{eq:14}\\
  E^\prime\!\!&=&\!\!\frac{m}{\sqrt{1-\frac{v^2}{c^2}}\,\sqrt{\frac{u^2}{c^2}-1}}(c^2-uv)\!<\!0,
  \;\;(u\!>\!\frac{c^2}{v}\;{\rm or}\; v\!>\!\frac{c^2}{u})\nonumber
\end{eqnarray}
Now the puzzle arises: How can a particle have $u^\prime<0\,(u>c^2/v)$ whereas its
$p^\prime>0$ ? How can it have energy $E^\prime<0$ whereas $E>0$ ? All of the above
puzzles from (\ref{eq:8})-(\ref{eq:14}) comprise the ``superluminal paradox".

The paradox disappears in a reasonable quantum theory (developed from the ``
reinterpretation principle") as follows: According to Bob's point of view, the tachyon
behaves in the $S^\prime$ frame (with $v>c^2/u$) just like  an antitachyon moving at a
velocity $u^\prime$. So its momentum and energy should be measured as:
\begin{equation}\label{eq:15}
  p^\prime_c=-p^\prime<0, \quad E^\prime_c=-E^\prime>0.
\end{equation}
This is because the well known operators in quantum mechanics:
\begin{equation}\label{eq:16}
  \hat{p}=-i\hbar\frac{\partial}{\partial x}, \quad \hat{E}=i\hbar\frac{\partial}{\partial t}
\end{equation}
are valid only for a particle. For its antiparticle, we should use instead:
\begin{equation}\label{eq:17}
  \hat{p}_c=i\hbar\frac{\partial}{\partial x}, \quad
  \hat{E}_c=-i\hbar\frac{\partial}{\partial t},
\end{equation}
(where the subscript $c$ refers to an antiparticle) which are just the essence of special
relativity (SR)\cite{11}.

Note that, however, the distinction between (\ref{eq:16}) and (\ref{eq:17}) is merely
relative, not absolute. For example, the energy of positron $e^+$ ( which is the
antiparticle of electron) in the process (\ref{eq:6}) is always positive like that of
neutron $n$. But once a neutrino has energy $E>0$ in the $S$ frame [see (\ref{eq:18})
below ] but has $E^\prime<0$ in the $S^\prime$ frame, it behaves just like an
antineutrino in the $S^\prime$ frame. Then the relations (\ref{eq:15}) and (\ref{eq:17})
must be taken into account when dealing with the LT. For further discussion, see the
Appendix.

\section{The threshold energy of protons and tachyon mass of neutrinos}

According to the present knowledge of particle physics, a free proton is stable in the
vacuum. It will never decay no matter how high its energy is (principle of relativity).
However, according to the theory of modern cosmology, low energy neutrinos including all
three flavors ($\nu_e,\;\nu_\mu$ and $\nu_\tau$) and  antineutrinos
($\bar{\nu}_e,\;\bar{\nu}_\mu$ and $\bar{\nu}_\tau$) are spreading through out space
isotropically. (There may also be considerable amount of high energy neutrinos directly
related to observable distant sources). So the process (\ref{eq:6}) could be induced for
either the subluminal or superluminal antineutrino if the proton has enough energy but
still well below $10^{15}$ eV, no knee will be seen. Hence, to explain the appearance of
a knee, which implies a sudden opening of a reaction channel, a new mechanism of
tachyonic neutrinos must be considered.

Now let us consider a process:
\begin{equation}\label{eq:18}
  \nu_e+p\rightarrow n+e^+,
\end{equation}
which  is strictly forbidden at energy $E_p<E_{th}$ due to the different lepton quantum
numbers on opposing sides. However, once the proton velocity $v$ exceeds a critical value
to be calculated below, a low energy neutrino $\nu_e$ (with $E_{\nu}\sim 0$ in the $S$
frame) suddenly transforms into an antineutrino with sufficiently high energy
$E^\prime_{\bar{\nu}}$ in the $S^\prime$ frame as discussed in the previous section. Then
the process (\ref{eq:6}) suddenly occurs as an exotic realization of (\ref{eq:18}) in the
S frame and contributes to the abrupt change of the slope in CRS as shown in
(\ref{eq:2}).

Denoting the rest masses ( and velocities in the $S^\prime$ frame) of protons, neutrons
and positrons by $m_p$, $m_n$ and $m_e$ ( $v^\prime_p$, $v^\prime_n$ and $v^\prime_e$ )
respectively, Bob can write down the conservation laws of energy and momentum  in the
$S^\prime$ frame as follows:
\begin{equation}\label{eq:19}
  m_p+\frac{m}{\sqrt{{u^\prime}^2/c^2-1}}
  =\frac{m_n}{\sqrt{1-{v^\prime}^2_n/c^2}}+\frac{m_e}{\sqrt{1-{v^\prime}^2_e/c^2}},
\end{equation}
\begin{equation}\label{eq:20}
  \frac{mu^\prime}{\sqrt{{u^\prime}^2/c^2-1}}=\frac{m_nv^\prime_n}{\sqrt{1-{v^\prime}^2_n/c^2}}
  +\frac{m_ev^\prime_n}{\sqrt{1-{v^\prime}^2_e/c^2}},
\end{equation}
where the antineutrino velocity $u^\prime$ is given by (\ref{eq:9}) with $v=v_p>0$ and
$u>0$. Any one of these velocities can take either positive [along the $x(x^\prime)$
axis] or negative [along the $-x(-x^\prime)$ axis] value automatically. We introduce
notations used in the theory of SR:
\begin{eqnarray}\label{eq:21}
  \beta_i&=&\frac{v_i}{c}=\tanh\zeta_i,\quad (i=p, n, e)\\\nonumber
  \gamma_{_i}&=&\frac{1}{\sqrt{1-\beta^2_i}}=\cosh\zeta_i,\quad
  \beta_i\gamma_i=\sinh\zeta_i,
\end{eqnarray}
where $\zeta_i$ is called the rapidity of particle $i$. Then (\ref{eq:19}) and
(\ref{eq:20}) read:
\begin{eqnarray}
  m_p+\frac{m}{\sqrt{{\beta^\prime}^2_\nu-1}}&=& m_n\cosh\zeta^\prime_n+m_e\cosh\zeta^\prime_e,\label{eq:22}\\
  m\frac{\beta^\prime_\nu}{\sqrt{{\beta^\prime}^2_\nu-1}}&=& m_n\sinh\zeta^\prime_n+m_e\sinh\zeta^\prime_e,\label{eq:23}
\end{eqnarray}
where $\beta^\prime_\nu=u^\prime/c<0$. Taking the square of (\ref{eq:22}) and
(\ref{eq:23}) respectively and substracting them each other, we get:
\begin{equation}\label{eq:24}
  \cosh(\zeta^\prime_n-\zeta^\prime_e)=\frac{1}{2m_n m_e}[m^2_p+\frac{2m_pm}{\sqrt{{\beta^\prime}^2_\nu-1}}
  -m^2_n-m^2_e-m^2].
\end{equation}
Since the hyperbolic function  $\cosh \xi\geq 1$, we find the condition for existence of
a solution to (\ref{eq:24}) being:
\begin{equation}\label{eq:25}
  \frac{2m_p m}{\sqrt{{\beta^\prime}^2_\nu-1}}>(m_n+m_e)^2+m^2-m^2_p,
\end{equation}
or
\begin{equation}\label{eq:26}
  \frac{1}{\sqrt{{\beta^\prime}^2_\nu-1}}>\frac{1}{\eta}\gg 1.
\end{equation}
Here a dimensionless parameter:
\begin{equation}\label{eq:27}
  \eta=\frac{2m_p m}{(m_n+m_e)^2+m^2-m^2_p}
\end{equation}
is defined. Rewriting (\ref{eq:9}) as ($\beta_\nu=u/c, \beta_p=v/c$):
\begin{equation}\label{eq:28}
  \beta^\prime_\nu=\frac{\beta_\nu-\beta_p}{1-\beta_\nu \beta_p}<0.
\end{equation}
we obtain the condition for the occurrence of process (\ref{eq:18}) in the $S$ frame as:
\begin{equation}\label{eq;29}
  \beta_p>\frac{\beta_\nu+\sqrt{1+\eta^2}}{\sqrt{1+\eta^2}\,\beta_\nu+1},
\end{equation}
\begin{equation}\label{eq:30}
  \frac{1}{\sqrt{1-\beta^2_p}}>\frac{\sqrt{1+\eta^2}\,\beta_\nu+1}{\eta\sqrt{\beta^2_\nu-1}}.
\end{equation}
Since $\eta\ll 1$, within a good approximation, we find the corresponding condition for
proton energy in the $S$ frame as ($c=1$):
\begin{equation}\label{eq:31}
  E_p=\frac{m_p}{\sqrt{1-\beta^2_p}}>\frac{m_p}{\eta}\sqrt{\frac{\beta_{_\nu}+1}{\beta_{_\nu}-1}}
  =\frac{m_p}{\eta}\sqrt{\frac{p_{_\nu}+E_\nu}{p_{_\nu}-E_\nu}}
\end{equation}
As discussed at the beginning of this section, the first knee in CRS should be the
threshold value of $E_p$ in (\ref{eq:31}) with $\beta_\nu\rightarrow \infty$ (i.e.,
$E_\nu\rightarrow 0$):
\begin{eqnarray}
  E_{th}&=&E^{(1)}_{th}=\frac{1}{2m}[(m_n+m_e)^2+m^2-m^2_p\,]\nonumber\\
  &\simeq &\frac{1.695\times 10^{15}}{m}\;{\rm eV}\label{eq:32}
\end{eqnarray}
where the value of tachyon mass of neutrino is in unit of eV/$c^2$. If we adopt the value
of $E^{(1)}_{th}=3.16\times 10^{15}$ eV as in Ref.~\cite{2}, we find the tachyon mass:
\begin{equation}\label{eq:33}
   m=m(\nu_e)=0.54\;\; {\rm eV}/c^2.
\end{equation}
Similarly, based on a known semileptonic decay mode of hyperon $\Lambda$ \cite{12}:
\begin{equation}
  \Lambda\rightarrow p+\mu^-+\tilde{\nu}_\mu,
\end{equation}
we may consider the process (\ref{eq:7}) induced in the $S^\prime$ frame right at the
threshold energy of the second knee in CRS, $E^{(2)}_{th}=6.31\times 10^{17}$ eV, for the
exotic reaction $\nu_\mu+p\rightarrow \Lambda+\mu^+$ triggered in the $S$ frame. Thus we
find ( $m_\Lambda=1115.6$ MeV/$c^2$, $m_\mu=105.7$ MeV/$c^2$):
\begin{eqnarray}
  E^{(2)}_{th}&=&\frac{m_p}{\eta^\prime}=\frac{1}{2m^\prime}[(m_\Lambda+m_\mu)^2+{m^\prime}^2-m^2_p\,]\nonumber\\
  &\simeq &\frac{3.056\times 10^{17}}{m^\prime}\;\;{\rm eV},\label{eq:34}
\end{eqnarray}
\begin{equation}\label{eq:35}
   m^\prime=m(\nu_\mu)=0.48\;\;{\rm eV}/c^2.
\end{equation}

\section{Neutrino oscillation and a minimal three-flavor model}

As shown in the particle table published in 2000 \cite{13}, the mass square of electron
neutrino defined by
\begin{equation}\label{eq:36}
  E^2=p\,c^2+m^2(\nu_e)c^4
\end{equation}
seems negative in tritium beta decay:
\begin{equation}\label{eq:37}
  m^2(\nu_e)=-2.5\pm 3.3\;\; {\rm eV}^2
\end{equation}
(see also \cite{7}). Due to difficulties in experiments and theoretical analysis
\cite{14}, physicists often think the present data is not accurate enough to fix the
value of $m(\nu_e)$. However, the first reason why its uncertainty is so large lies in
the fact that a neutrino is oscillating among three flavors as verified by the Kamiokande
\cite{15} and SNO \cite{16} experimental groups. The oscillation implies that neutrinos
are staying (at least) in two mass eigenstates. For instance, in a minimal three-flavor
model for tachyonic neutrino \cite{17}, an equation containing only one parameter
$\delta$ is proposed ($\hbar=c=1$):
\begin{equation}\label{eq:38}
  \left\{\begin{array}{rcl}
    i\dot{\xi}_e&=&i\vec{\sigma}\cdot\bigtriangledown\xi_e-\delta(\eta_\mu+\eta_\tau)\\
    i\dot{\eta}_e&=&-i\vec{\sigma}\cdot\bigtriangledown\eta_e+\delta(\xi_\mu+\xi_\tau)\\
    i\dot{\xi}_\mu&=&i\vec{\sigma}\cdot\bigtriangledown\xi_\mu-\delta(\eta_\tau+\eta_e)\\
    i\dot{\eta}_\mu&=&-i\vec{\sigma}\cdot\bigtriangledown\eta_\mu+\delta(\xi_\tau+\xi_e)\\
    i\dot{\xi}_\tau&=&i\vec{\sigma}\cdot\bigtriangledown\xi_\tau-\delta(\eta_e+\eta_\mu)\\
    i\dot{\eta}_\tau&=&-i\vec{\sigma}\cdot\bigtriangledown\eta_\tau+\delta(\xi_e+\xi_\mu)\\
  \end{array}\right.
\end{equation}
where $\xi_i (i=e, \mu, \tau)$ and $\eta_i$ are the left-handed and right-handed chiral
states of flavor $i$ for a neutrino ( $\vec{\sigma}$ are Pauli matrices). The neutrino is
oscillating among three mass eigenstates of energy square being :
\begin{equation}\label{eq:39}
    E_j=p^2-m^2_j,\quad (j = 1,2,3)
\end{equation}
\begin{equation}\label{eq:40}
    m^2_1=4\delta^2,\quad m^2_2=m^2_3=\delta^2.
\end{equation}
In this model, however, different flavors all have the same mass. The fitting values of
$\nu_e$ and $\nu_\mu$ from (\ref{eq:33}) and (\ref{eq:35}) seem to favor the predictions
above. And the large uncertainty in (\ref{eq:37}) is primarily due to the existence of
oscillation between two mass eigenvalues:
\begin{equation}\label{eq:41}
   m_1=2\delta,\quad  m_2(=m_3)=\delta.
\end{equation}
As discussed in \cite{17}, the expectation value of mass square of $\nu_e$ just created
from beta decay should be:
\begin{equation}\label{eq:42}
   m^2(\nu_e)=-\frac{8}{5}\delta^2\pm\frac{6}{5}\delta^2
\end{equation}
in comparison with (\ref{eq:37}). On the other hand, the two knees in CRS should be
fitted by one expectation value of tachyon mass for a neutrino in flight:
\begin{equation}\label{eq:43}
  \widetilde{m}(\nu_e)=\widetilde{m}(\nu_\mu)=\frac{3}{2}\delta\pm \frac{1}{2}\delta.
\end{equation}
Comparing it with the average of (\ref{eq:33}) and (\ref{eq:35}), 0.51 eV/$c^2$, we find
the value of $\delta$ being approximately:
\begin{equation}\label{eq:44}
  \delta=0.34\;\; {\rm eV}.
\end{equation}

However, we suggest that if experimental physicists can treat their data in a two-center
fitting as shown by (\ref{eq:40}) or (\ref{eq:41}) in $4:1$ or $2:1$ ratio (with
statistical weight ratio $1:4$ or $1:1$) for the case of (\ref{eq:42}) or (\ref{eq:43})
respectively, better results could be obtained.

\section{Summary and discussion}

(a) Following Kostelecky and Ehrlich, we elaborate a model of tachyonic neutrinos to
explain two knees in the CRS by two Eqs.~(\ref{eq:32}) and (\ref{eq:34}) respectively
with one tachyon mass of neutrino ($\nu_e$ or $\nu_\mu$) being estimated to be around
0.51 eV/$c^2$ regardless of the flavor. This result favors a minimal three-flavor model
for tachyonic neutrinos containing only one coupling parameter $\delta=0.34$ eV.

(b) The tachyon theory for neutrinos is a natural extension of the theory of special
relativity (SR) in combination with the theory of quantum mechanics (QM). Especially, the
basic operator relations (\ref{eq:16}) for particles should be supplemented by
(\ref{eq:17}) for antiparticles while the addition law for velocities in SR,
Eq.~(\ref{eq:9}), remains valid for both subluminal and superluminal motions.

(c) We dare not discuss the explanation of the ankle in CRS before further calculation
could be made in detail which might be related to the distribution of distant sources for
both cosmic rays and neutrinos. However, one thing can supplement the interpretation of
the evasion of GZK cutoff \cite{3, 4, 5, 6, 7}. In the long chain of decays:
$p\rightarrow n\rightarrow p\rightarrow n\rightarrow\cdots$ (or $p\rightarrow
\Lambda\rightarrow p\rightarrow \Lambda\rightarrow\cdots$), the lifetime of $n$ (or
$\Lambda$) may be different for different polarizations: while the right-handed $n$
($\Lambda$) has lifetime $\tau_{_R}$, the left-handed one has $\tau_{_L}$ \cite{20, 21}:
\begin{equation}\label{eq:45}
  \tau_{_R}=\frac{\tau}{1-\beta},\quad \tau_{_L}=\frac{\tau}{1+\beta},
\end{equation}
where $\tau=\tau_0/\sqrt{1-\beta^2}$ ($\beta=v/c$). The faster the speed $v$ of $n$
($\Lambda$) is, the larger $\tau_{_R}$ will be. Therefore, we expect that most nucleons
(fermions) in the cosmic ray should be right-handed polarized. Future experiments will
pose a serious test on the mechanism of the knees, ankle and the evasion of GZK cutoff as
well as the prediction of (\ref{eq:45})---a phenomenon of parity violation in the beta
decay.

(d) A particle is always impure in the sense of having two contradictory fields,
$\varphi$ and $\chi$. They obey the basic symmetry of space-time inversion ($x\rightarrow
-x ,\; t\rightarrow -t$):
\begin{equation}\label{eq:46}
  \varphi(-x,-t)\rightarrow \chi(x,t),\quad \chi(-x,-t)\rightarrow \varphi(x,t).
\end{equation}
The neutrino is no exception to this rule. Indeed, Eq.~(\ref{eq:38}) with relations:
\begin{equation}\label{eq:47}
  \varphi_i=\frac{1}{\sqrt{2}}(\xi_i+\eta_i),\quad \chi_i=\frac{1}{\sqrt{2}}(\xi_i-\eta_i)
\end{equation}
remains invariant under the transformation (\ref{eq:46}) with subscripts added. However,
uniquely, neutrinos have another two symmetries: Eq.~(\ref{eq:38}) is also invariant
under the pure time inversion ($x\rightarrow x,\; t\rightarrow -t$):
\begin{equation}\label{eq:48}
  \xi_i(x,-t)\rightarrow \eta_i(x,t),\quad \eta_i(x,-t)\rightarrow \xi_i(x,t).
\end{equation}
This is why Bob will see a neutrino (in the $S$ frame) transforming into an antineutrino
in the $S^\prime$ frame. On the other hand, the left-right symmetry (parity) is violated
to maximum, since Eq.~(\ref{eq:38}) is no longer invariant under the pure space inversion
( $x\rightarrow -x,\; t\rightarrow t$):
\begin{equation}\label{eq:49}
    \xi_i(-x,t)\rightarrow \eta_i(x,t),\quad \eta_i(-x,t)\rightarrow \xi_i(x,t).
\end{equation}
in contrast to the case of the Dirac equation. Interestingly enough, a massive neutrino
(antineutrino) can preserve its permanent left-handed (right-handed) polarization because
its velocity $u$ exceeds the speed of light $c$.

(e) The mass $m$ and energy $E$ of every particle or antiparticle (regardless of it being
a tachyon or not) is real and positive in the strict sense that they are measured in
certain experiments. However, for a theory capable of treating the particle and
antiparticle on an equal footing, it must be invariant under symmetry transformation:
$m\rightarrow -m$ \cite{18}. Hence, from the theoretical point of view, by using
Eq.~(\ref{eq:16}) only, we may say that the antiparticle state is the negative-energy
state of a particle, e.g., the wavefunction (\ref{eq:a12}) versus (\ref{eq:a11}) in the
Appendix. There is an interesting question relevant to the theory of cosmology: Where the
enormous energy of our universe (comprising mainly of matter) comes from? In other words,
is the energy conserved in the bigbang? We try to answer the above question by assuming
that during the bigbang, equal amounts of matter and antimatter were created
simultaneously \cite{18}. Hence, in some sense, we may regard the entire universe really
as an ultimate free lunch \cite{19}.

\begin{acknowledgments}
I am greatly indebted to P. T. Leung for bringing Ref.~[7] to my attention. I also thank
S. Q. Chen and Z. Q. Shi for helpful discussions.
\end{acknowledgments}

\appendix*
\section{A tachyonic neutrino as a microscopic Schr\"{o}dinger's cat}

Bob never observed in the $S^\prime$ frame a neutrino moving backward in time (because
``time" is a conception endowed by the observer, not by the particle) but an antineutrino
flying toward him with high energy $E^\prime_{\tilde{\nu}}>0$ and momentum
$p^\prime_{\tilde{\nu}}=\beta^\prime_\nu E^\prime_{\tilde{\nu}}<0$. So he writes down the
conservation laws of energy and momentum in the form of (\ref{eq:19}) and (\ref{eq:20}),
i.e.,
\begin{eqnarray}
    E^\prime_p+ E^\prime_{\tilde{\nu}}&=&E^\prime_n+E^\prime_e, \label{eq:a1}\\
    p^\prime_{\tilde{\nu}}&=&p^\prime_n+p^\prime_e.\label{eq:a2}
\end{eqnarray}
If Alice wishes to find corresponding laws in the $S$ frame, she first resorts to the LT
[like (\ref{eq:11})] for the neutron and positron, yielding:
\begin{eqnarray}
  E_n+E_e&=&\frac{(E^\prime_n+E^\prime_e)+\beta_p(p^\prime_n+p^\prime_e)}{\sqrt{1-\beta^2_p}},\label{eq:a3}\\
  p_n+p_e&=&\frac{(p^\prime_n+p^\prime_e)+\beta_p(E^\prime_n+E^\prime_e)}{\sqrt{1-\beta^2_p}}.\label{eq:a4}
\end{eqnarray}
Substituting (\ref{eq:a1}) and (\ref{eq:a2}) into the right sides of (\ref{eq:a3}) and
(\ref{eq:a4}) and noticing from (\ref{eq:28}) that
\begin{equation}\label{eq:a5}
  \frac{1}{\sqrt{{\beta^\prime}^2_\nu-1}}=\frac{\beta_\nu\beta_p-1}{\sqrt{\beta^2_\nu-1}\;
  \sqrt{1-\beta^2_p}}>0,
\end{equation}
\begin{equation}\label{eq:a6}
  E^\prime_{\tilde{\nu}}=\frac{p_\nu\beta_p-E_\nu}{\sqrt{1-\beta^2_p}}>0,\quad
  p^\prime_{\tilde{\nu}}=\frac{E_\nu\beta_p-p_\nu}{\sqrt{1-\beta^2_p}}<0.
\end{equation}
Alice finds from (\ref{eq:a3}) and (\ref{eq:a4}) that ($E_\nu>0,\;p_\nu=\beta_\nu E_\nu>
0$):
\begin{eqnarray}
    E_n+E_e&=&E_p-E_\nu,\label{eq:a7}\\
    p_n+p_e&=&p_p-p_\nu.\label{eq:a8}
\end{eqnarray}
At first sight, (\ref{eq:a7}) and (\ref{eq:a8}) are very strange for the process
(\ref{eq:18}), but they are inevitable since both A and B insist on treating every
particle (antiparticle) having positive energy and momentum along its velocity. Both A
and B believe in the LT for linking the quantities they measured. Thus A and B come to
agree that the minus sign before the $E_\nu$ and $p_\nu$ is due to the fact that the same
particle exhibits itself as a neutrino in the $S$ frame whereas it must appear as an
antineutrino in the $S^\prime$ frame before it can be absorbed by the proton.

If instead of (\ref{eq:a7}) and (\ref{eq:a8}), Alice insists on writing conservation laws
in the $S$ frame as usual:
\begin{eqnarray}
    E_n+E_e&=&E_p+E_\nu,\label{eq:a9}\\
    p_n+p_e&=&p_p+p_\nu.\label{eq:a10}
\end{eqnarray}
Then after measurements and calculation, Alice has to admit that both $E_\nu$ and $p_\nu$
here turn to negative values as shown by (\ref{eq:a7}) and (\ref{eq:a8}). When doing so,
Alice is tacitly assuming that before the neutrino interacts with the proton, $E_\nu (>
0)$ and $p_\nu (>0)$ are already existing. Hence she is worried that the energy (momentum
) conservation law seems to be violated in the process (\ref{eq:18}).

The above discussion reminds us of a remarkable experiment on SQUID \cite{22}, showing a
macroscopic Schr\"{o}dinger's cat puzzle---in a superconducting ring carrying clockwise
current, the ``hidden" anticlockwise current was measured by the absorption of microwave
radiation. (For its discussion, see \cite{23}, also \cite{10}).

Now a tachyonic neutrino is also a Schr\"{o}dinger's cat but on a microscopic scale. As
shown in (\ref{eq:38}), a plane wavefunction (WF) contains all six fields $\xi_i$ and
$\eta_i (i=e, \mu, \tau)$:
\begin{equation}\label{eq:a11}
  \xi_i\sim\eta_i\sim exp\,[\frac{i}{\hbar}\,(p_\nu\, x-E_\nu\, t)],\quad (|\xi_i|>|\eta_i|)
\end{equation}
and describes a neutrino with 100 \% left-handed polarization. We may regard $\xi_i$
being the ``alive-cat" state and $\eta_i$ the ``dead-cat" state. The latter is in a
subordinate status and so doesn't appear as an explicit right-handed antineutrino
ingredient. But once the neutrino is absorbed by a proton, the ``hidden" $\eta_i$ is
suddenly activated and dominates the $\xi_i$---together they show up as an antineutrino
with 100 \% right-handed polarization in the $S^\prime$ frame as described by the WF :
\begin{equation}\label{eq:a12}
  \eta^\prime_i\sim\xi^\prime_i\sim exp\,[-\frac{i}{\hbar}(p^\prime_{\tilde{\nu}}\, x^\prime-E^\prime_{\tilde{\nu}}\, t^\prime)],
  \quad (|\eta^\prime_i|>|\xi^\prime_i|)
\end{equation}
In our point of view, the momentum $p^\prime_{\tilde{\nu}}$ and energy
$E^\prime_{\tilde{\nu}}$ do not exist until the antineutrino is absorbed by the proton.
Similarly, before the neutrino is absorbed in the $S$ frame, the values of $p_\nu$ and
$E_\nu$ also do not exist. We need not worry about conservation laws like (\ref{eq:a9})
and (\ref{eq:a10}) since all quantities written there are absent until created during the
occurrence of process (\ref{eq:18}). We should not interpret the quantum state and WF too
materially. The WF is merely a probability amplitude of ``fictitious measurement", not
real measurement.

A similar situation happened in another remarkable Which-Way (WW) experiment performed on
an atom interferometer \cite{24,25}. Some physicists were worried that the uncertainty
relation
\begin{equation}\label{eq:a13}
  \triangle p\;\triangle x\geq\hbar/2
\end{equation}
might be invalid for the momentum $p$ and position $x$ of the atom's center-of-mass. But
actually, $p$ and $x$ had not been measured in the motion of atom and so
Eq.~(\ref{eq:a13}) has nothing to do with the WW experiment \cite{26}. What is measured
is the WW information of atomic internal states gained from the absorption of microwave
pulses. In fact, the authors themselves already correctly wrote down a complementary
relation for the distinguishability (of WW information) $D$ and the fringe visibility $V$
as \cite{25}:
\begin{equation}\label{eq:a14}
   D^2+V^2\leq 1.
\end{equation}

For further discussion, see \cite{10}.

\end{document}